\documentclass[sigconf, nonacm]{acmart}

\setcopyright{none}
\renewcommand\footnotetextcopyrightpermission[1]{} 
\AtBeginDocument{%
  }

\usepackage{algorithm}
\usepackage{algpseudocode}
\usepackage{graphicx}
\usepackage{subcaption}
\usepackage{tikz}
\usetikzlibrary{positioning, shapes.geometric}
\usepackage{float}
\usepackage{svg}
\usepackage{caption}
\usepackage{tablefootnote}
\usepackage{placeins}
\usepackage{dblfloatfix}
\begin{document}
\title{Transforming Location Retrieval at Airbnb: A Journey from Heuristics to Reinforcement Learning}

\author{Dillon Davis}
\email{dillon.davis@airbnb.com}
\orcid{1234-5678-9012}
\affiliation{%
  \institution{Relevance}
  \institution{Airbnb, Inc.}
  \city{San Francisco, CA}
  \country{USA}
}
\author{Huiji Gao}
\email{huiji.gao@airbnb.com}
\affiliation{%
  \institution{Relevance}
  \institution{Airbnb, Inc.}
  \city{San Francisco, CA}
  \country{USA}
}

\author{Thomas Legrand}
\email{thomas.legrand@airbnb.com}
\affiliation{%
  \institution{Relevance}
  \institution{Airbnb, Inc.}
  \city{San Francisco, CA}
  \country{USA}
}

\author{Weiwei Guo}
\email{weiwei.guo@airbnb.com}
\affiliation{%
  \institution{Relevance}
  \institution{Airbnb, Inc.}
  \city{San Francisco, CA}
  \country{USA}
}

\author{Malay Haldar}
\email{malay.haldar@airbnb.com}
\affiliation{%
  \institution{Relevance}
  \institution{Airbnb, Inc.}
  \city{San Francisco, CA}
  \country{USA}
}
\author{Alex Deng}
\email{alex.deng@airbnb.com}
\affiliation{%
  \institution{Relevance}
  \institution{Airbnb, Inc.}
  \city{San Francisco, CA}
  \country{USA}
}
\author{Han Zhao}
\email{han.zhao@airbnb.com}
\affiliation{%
  \institution{Relevance}
  \institution{Airbnb, Inc.}
  \city{San Francisco, CA}
  \country{USA}
}
\author{Liwei He}
\email{liwei.he@airbnb.com}
\affiliation{%
  \institution{Relevance}
  \institution{Airbnb, Inc.}
  \city{San Francisco, CA}
  \country{USA}
}
\author{Sanjeev Katariya}
\email{sanjeev.katariya@airbnb.com}
\affiliation{%
  \institution{Relevance}
  \institution{Airbnb, Inc.}
  \city{San Francisco, CA}
  \country{USA}
}

\renewcommand{\shortauthors}{Dillon Davis et al.}
\begin{abstract}
The Airbnb search system grapples with many unique challenges as it continues to evolve. We oversee a marketplace that is nuanced by geography, diversity of homes, and guests with a variety of preferences. Crafting an efficient search system that can accommodate diverse guest needs, while showcasing relevant homes lies at the heart of Airbnb's success. Airbnb search has many challenges that parallel other recommendation and search systems but it has a unique information retrieval problem, upstream of ranking, called location retrieval. It requires defining a topological map area that is relevant to the searched query for homes listing retrieval. The purpose of this paper is to demonstrate the methodology, challenges, and impact of building a machine learning based location retrieval product from the ground up. Despite the lack of suitable, prevalent machine learning based approaches, we tackle cold start, generalization, differentiation and algorithmic bias. We detail the efficacy of heuristics, statistics, machine learning, and reinforcement learning approaches to solve these challenges, particularly for systems that are often unexplored by current literature.
\end{abstract}

\ccsdesc[500]{Information systems~Recommender systems}
\ccsdesc[500]{Information systems~Novelty in information retrieval}
\ccsdesc[500]{Information systems~Query intent}
\ccsdesc[500]{Applied computing~Online shopping}
\keywords{Information Retrieval, Audience Expansion, Location based search systems, Machine Learning, Deep Learning, Reinforcement Learning, E-commerce}


\maketitle

\section{Introduction}
\begin{figure}[h]
  \centering
  \includegraphics[width=0.75\linewidth]{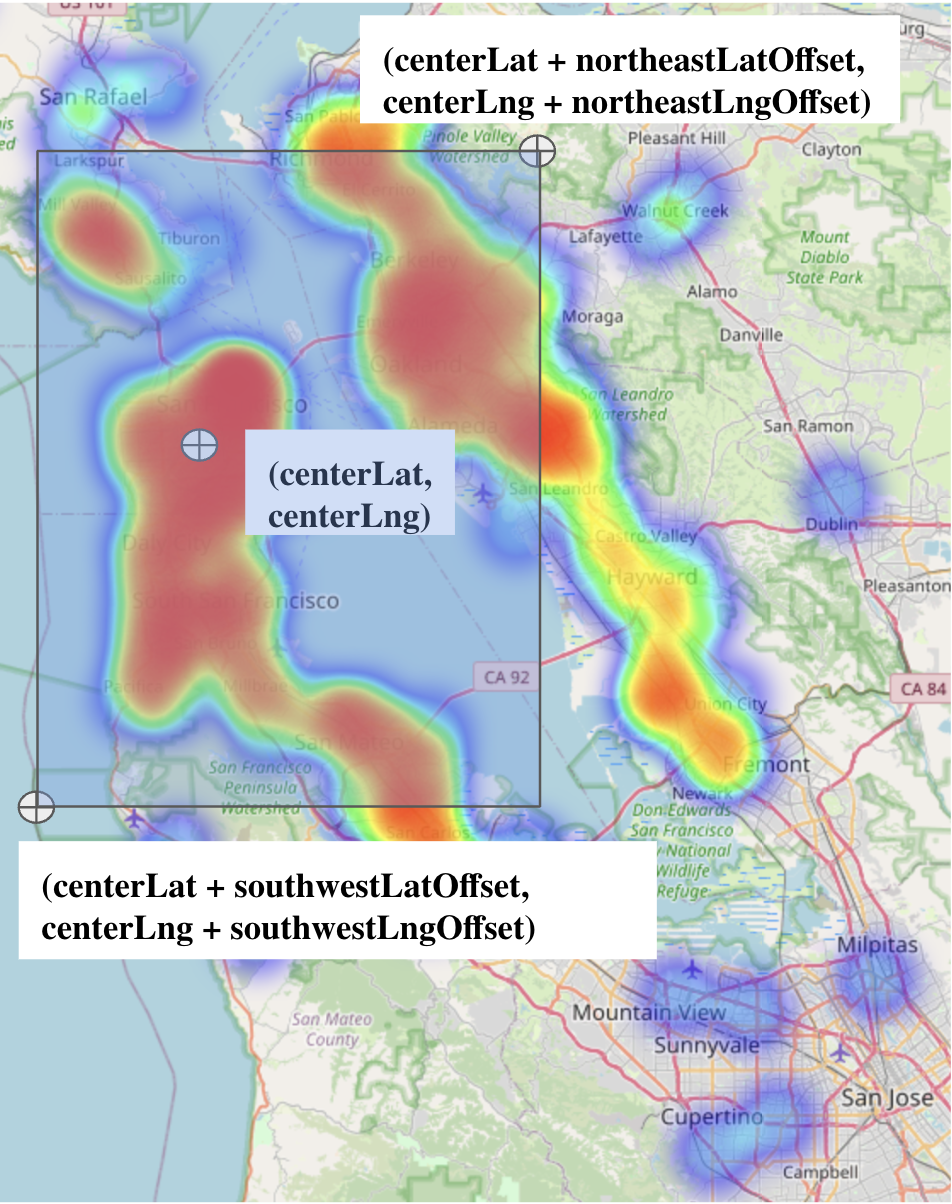}
  \caption{Heatmap of booked listing locations for guests who searched for "San Francisco, CA"}
  \label{sf_heatmap}
\end{figure}

    \quad Airbnb is a trusted platform for hosts with diverse homes everywhere around the world. It serves guests with increasingly unique preferences about location, amenities, style, price, and more. 
Traditional travel accommodations are often concentrated in a few limited areas of a searched location. However, Airbnb often has inventory spread throughout the surrounding areas with varying density, amenities, styles, prices, space, and environments. For example, a family who may not like the limited inventory available in a city with short term rental regulations, typically might stay at a hotel for a trip. However, single family home inventory in cities right outside of the regulated city may suit them much better than the semiprivate, smaller accommodations in the city. This is evidenced by the extensive booking behavior of San Francisco searchers in surrounding areas like Daly City, shown in Figure \ref{sf_heatmap}. \\
    \quad This presents Airbnb the opportunity to truly allow any guest to belong anywhere, regardless of their needs. We can infer where guests \textbf{might} be willing to stay for any given destination and expose different kinds of inventory available across those destinations. If we can do this, while respecting their flexibility and preferences, we can provide guests with much better travel experiences and appeal to a wider audience. \\
\null    \quad A typical guest journey starts by entering a destination into the search bar (a location search) and receiving a list of the top ranked, most bookable results for their search. Some guests ($\sim$52\%) then pan and zoom the displayed map (a map search) to find more listings that may be more suitable for them. Many guests never go beyond the initial location search to explore Airbnb's diverse inventory in more detail. Exposing this unique, differentiated, bookable inventory in the initial location search can help all guests better understand the breadth and depth of Airbnb's offerings. \\
\null    \quad Airbnb has incredible scale with over 7M+ active listings, $\sim$390M nights booked per year and proportionately large search volume. It also has complex ranking systems with strict performance and product expectations. As a result, it is infeasible to rank \textbf{every listing} for \textbf{every search}. We also cannot simply filter by whether a listing matches the exact searched location, due to the value and diverse amenities of listings in surrounding areas discussed above. Our search system requires defining a rectangular area on the map, known as \textbf{\textit{retrieval bounds}}. These retrieval bounds are used to filter our inventory to a reasonable set of listings that any guest would \textbf{ever} book for the search. The search ranking models described in \cite{abdool2020managing} \cite{haldar2019applying} \cite{haldar2020improving}  \cite{tan2023optimizing} \cite{haldar2023learning} then determine the most relevant and bookable listings in these retrieval bounds, which are shown to Airbnb's guests. This flow is outlined below in Figure \ref{system_architecture} \\
 \begin{figure*}[h]
  \centering
  \includegraphics[width=0.75\linewidth]{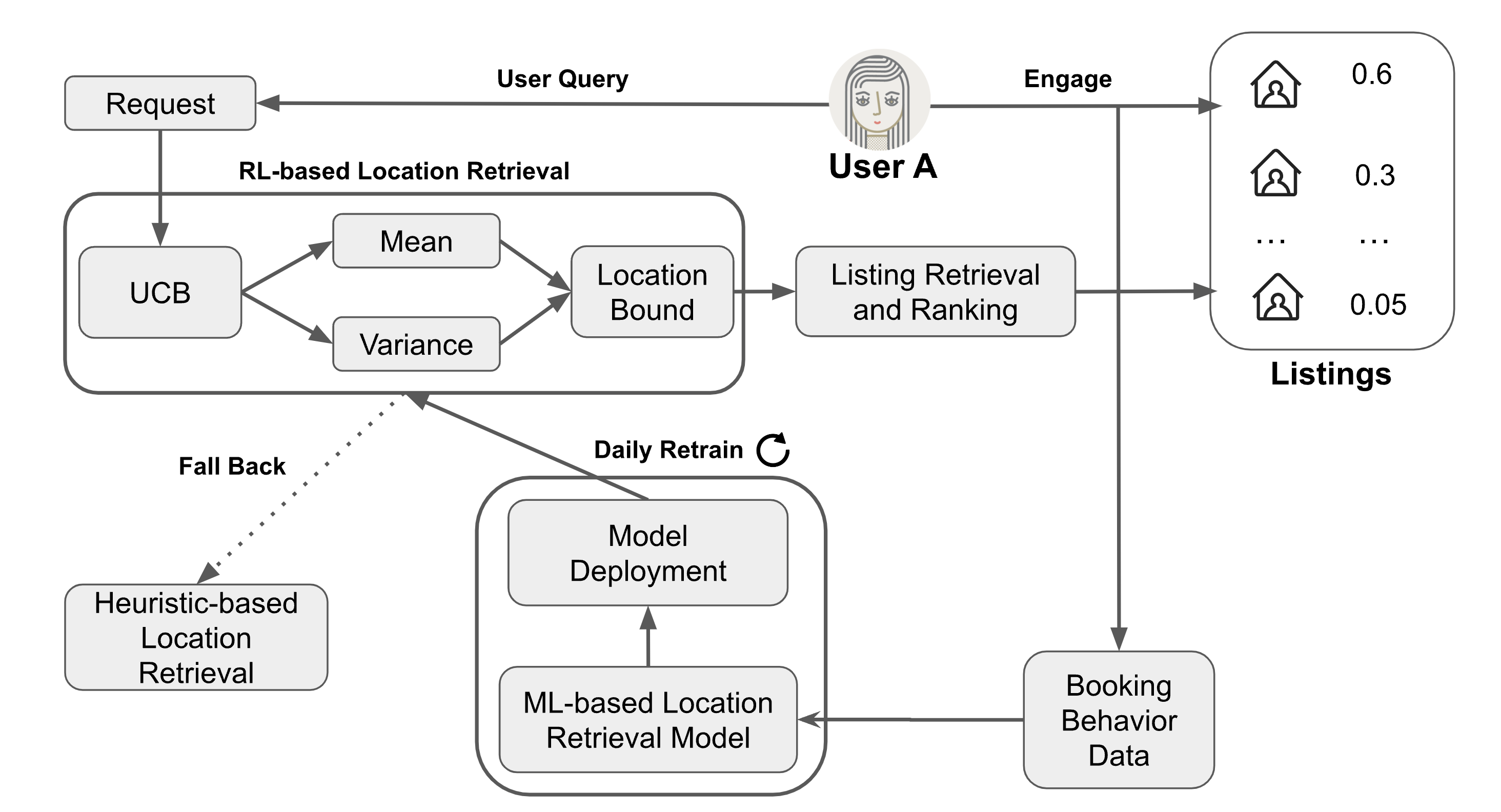}
  \caption{Location Retrieval System Architecture}
  \label{system_architecture}
\end{figure*} \\
\null    \quad These retrieval bounds are set by the guest for map searches but they are inferred for location searches. This retrieval bound inference based on the location and other search parameters is known as \textit{\textbf{location retrieval}}. It allows us to include the bookable Daly City inventory, demonstrated in Figure \ref{sf_heatmap}, for a San Francisco search, albeit typically ranked much lower since it does not strictly respect the guest's search criteria. Figure \ref{sf_heatmap} also shows the inferred retrieval bounds in blue for a sample San Francisco location search. \\

\null    \quad Location retrieval trades off increasing the size, quality, and diversity of the final result set for a search, at the cost of latency and potential location irrelevance. For example, narrow retrieval bounds can eliminate listings that might be relevant to a guest. On the other hand, wide retrieval bounds can include irrelevant listings and slow down search. \\
\null    \quad This presented the opportunity of building a data and machine learning driven product solution from the ground up to deeply understand this trade off by capturing booking preferences. However, understanding these preferences requires developing an end-to-end solution that collects, models, and enhances booking behavior data, by tackling the following unique challenges:
\begin{enumerate}
    \item \textbf{Cold Start}: How can we develop an intuitive, initial solution with little prior data while enabling data collection?
    \item  \textbf{Generalization and Differentiation}: How can we use prior data and bookings to build an ML solution that can generalize across searches and differentiate experiences?
\item    \textbf{Reinforcement}: How can we enhance data collected from bookings to build a robust, unbiased, high confidence model?
\end{enumerate}
  \quad There is little existing work on how to tackle the location retrieval problem even though location based search is used across the web every day on products like Yelp\texttrademark, Google Maps\texttrademark, Booking.com\texttrademark, and Zillow\texttrademark. The closest task is audience expansion for applications like marketing and ads which have been explored by many \cite{10.1145/2939672.2939680} \cite{10.1145/3394486.3403295} \cite{10.1145/3447548.3467179}. However, none of their approaches could directly be applied due to the large differences in task construction and output such as optimizing similarity instead of diversity and product experience constraints regarding the distance of listings. This paper will go over the evolution and efficacy of solutions for the outlined challenges, including simple heuristics, statistics, machine learning and reinforcement learning, with an emphasis on the application of reinforcement learning.
\section{Cold Start Heuristics for Location Retrieval}
\null \quad The initial approach for location retrieval was simple heuristics based on the type of location that was searched, i.e., country, state, city, neighborhood, address, etc. They typically took advantage of the administrative bounds that outline the exact map area corresponding to the searched location. The heuristics for each location type are outlined below.
\begin{enumerate}
	\item \textbf{Countries, states, neighborhoods (Heuristic 1)}: Use the administrative bounds from geocoding services to retrieve listings that are exactly within the searched location.
	\item \textbf{Cities (Heuristic 2)}: We construct retrieval bounds with a 25 mile radius around the searched location’s center
	\item \textbf{Addresses and buildings (Heuristic 3)}: We used retrieval bounds constructed by scaling the administrative bounds from geocoding services by 2.5X to prevent showing little to no results and better match booking behavior.
\end{enumerate}
\vspace{-4mm}
\null \quad These heuristics served as simple cold start techniques that allowed the product to function as intended as well as collect information over time about booking  preferences. We also improved \textbf{Heuristic 3} by introducing a parameterized smooth function that computes an expansion factor for the original diagonal size of the administrative bounds \textbf{(Heuristic 4)}. The log scale function \begin{math} f \end{math} scaled the original diagonal size by a hyperparameter \begin{math} \mathbf{\beta} \end{math} and offset \begin{math} \mathbf{\alpha} \end{math} shown below. This function better captured inventory in the surrounding areas of these very specific locations, providing guests with more options.
\begin{equation}
    f(d) = max(1.0, \alpha + \beta * ln(d + 1)).
\end{equation}
where d is the administrative bounds diagonal size, \begin{math}\alpha = 2.9 \end{math} and \begin{math}\beta = -0.5\end{math}. Hyperparameter values were chosen by maximizing expansion while still respecting the performance constraints of our search system.
\section{Statistics based Location Retrieval}
\vspace{-4mm}
\null \quad After solving cold start, we slowly collected millions of examples of bookings indicating where a guest is willing to stay when they search for a particular location. However, the naive cold start solutions often returned few, unbookable search results. This led us to ask: what’s the simplest way we can use this data to deliver a search experience that’s more aligned with guest booking behavior? \\
\null \quad The simplest solution was to build a dataset for each travel destination, with locations of listings that are booked when guests search for that destination. We then constructed retrieval bounds that contained 96\% of the booked listing locations that were closest to the center of a given searched location, and applied a small expansion factor on top of that. This was selected to maximize expansion while still respecting the performance constraints of our search system. These retrieval bounds were simply keyed by the searched location and were not differentiated based on other search parameters, such as trip dates and number of guests. \\
\null \quad We believed that this solution would be much more robust than the naive heuristics above because it aligned better with guest's booking behavior in a simple intuitive manner. 
\section{Machine Learning based Location Retrieval}
\subsection{The opportunity}

  \begin{figure}[h]
   
        \includegraphics[width=0.75\linewidth]{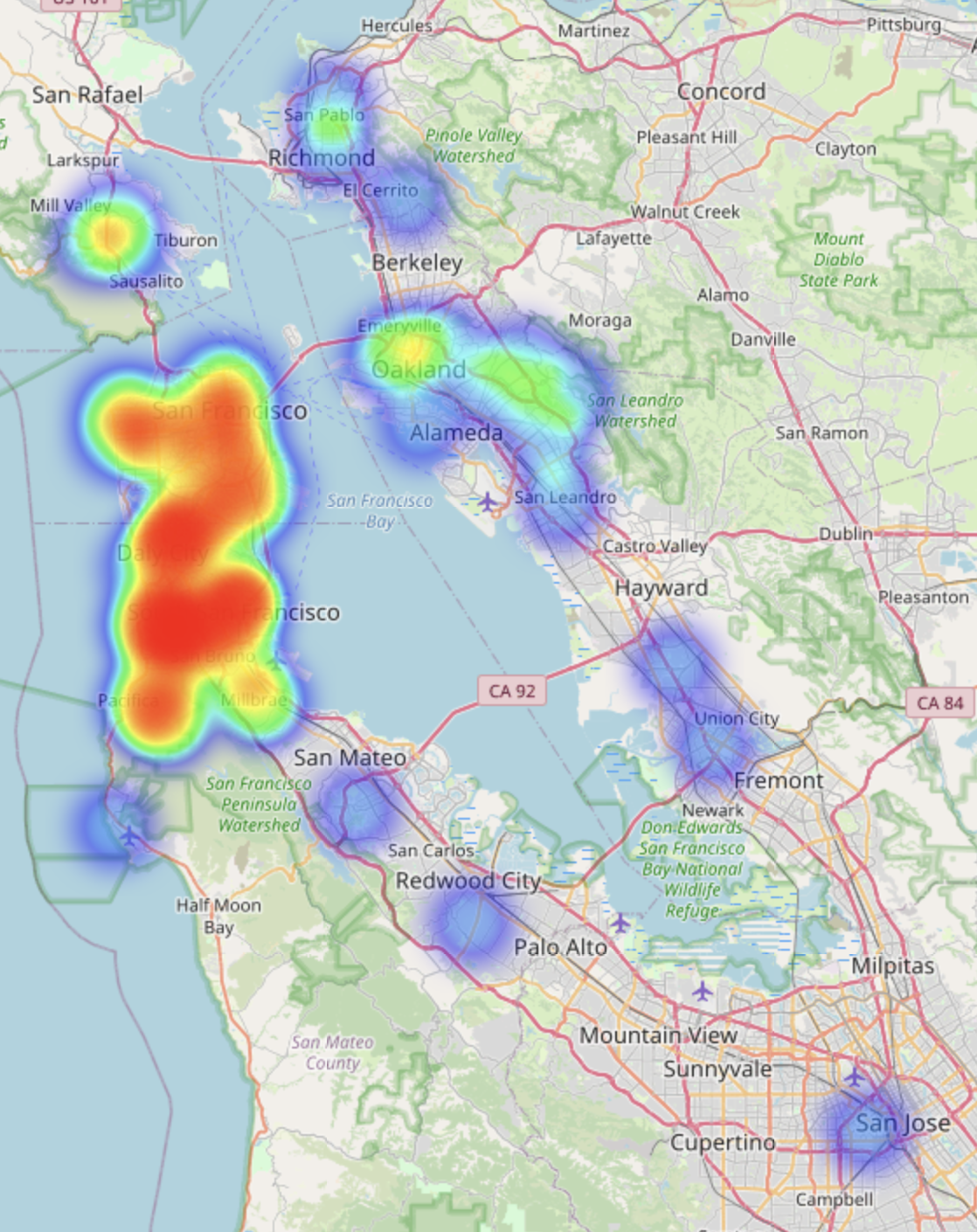}
  \caption{Heatmap of booked listing locations for guests who searched for "San Francisco, CA" with guest count >= 10}
  \label{sf_heatmap_group}
  \end{figure}
\vspace{-4mm}
\null \quad Beyond naively applying where Airbnb's guests had booked in the past for a location, we believed that there was an opportunity to \textbf{learn} from this data. This could generalize better for locations with little data, differentiate search experiences based on specific search parameters, and improve the experiences of non-bookers by learning from booker behavior. \\
\null \quad For example, large group travelers searching for San Francisco book in significantly different locations than San Francisco searchers overall, demonstrated in Figure \ref{sf_heatmap} and \ref{sf_heatmap_group}. Differentiating search experiences based on parameters like guest count and lead time could potentially provide more succinct, relevant results for guests.
\subsection{The Approach}
\vspace{-4mm}
\null \quad The location retrieval problem can be formulated in the following way in order to generalize and differentiate searches based on prior data and search parameters. We define the location retrieval model as a function \begin{math} \mathbf{f} \end{math}, that takes features derived from a search request \begin{math} \mathbf{x} \end{math} and produces a location bound \begin{math} \mathbf{y} \end{math}, which contains 4 floats 
\begin{equation}
\begin{split}
\mathbf{y} = [swLatOffset, neLatOffset, \\
swLngOffset, neLngOffset]
\end{split}
\end{equation}
where \begin{math}\mathbf{swLatOffset}, \mathbf{neLatOffset},\end{math} \begin{math} \mathbf{swLngOffset}, \mathbf{neLngOffset} \end{math} correspond to the southwest latitude offset, northeast latitude offset, southwest longitude offset, and northeast longitude offset in degrees. These offsets can be combined with the center latitude and center longitude of the searched location (\begin{math}\mathbf{centerLat}\end{math} and \begin{math}\mathbf{centerLng}\end{math}) to construct the predicted retrieval bounds (\begin{math}\mathbf{retrievalSwLat}, \end{math} \\
\begin{math}\mathbf{retrievalSwLng},\mathbf{retrievalNeLat},\mathbf{retrievalNeLng}\end{math}), illustrated in Figure \ref{sf_heatmap}. \\
\null \quad Location retrieval has opposing objectives of including listings that anyone would ever book (recall maximization) while minimizing the size of the retrieval bounds for performance. These opposing objectives did not lend themselves well to traditional machine learning model formulations and losses. Given the nature of retrieval bounds based retrieval, there are no true negatives with respect to location retrieval when a guest books a listing (a positive). If a guest searches for San Francisco and books a listing in Oakland, that does not indicate that we should not have retrieved listings from San Francisco. \\
\null \quad Considering these challenges, we took the following approach for the model formulation.
\begin{itemize}
    \item \textbf{Training Examples}:
    We use 1 year of training examples ($\sim$365M) derived from bookings and searches. We attribute any booking that is not cancelled after 7 days to any search issued on the same day or 1 day before the reservation is made. The booked listing must also appear in the attributed search's results.
    \item \textbf{Training Features}: We employ features derived directly from the search request. These features are a mix of 9 continuous features that are directly used in the model and 10 categorical features that are embedded during training time and trained in unison with the model. The most important features based on the impact of randomly permuting each feature on booked listing location recall  are following in decreasing order. Five unimportant features are omitted for company privacy.
    \begin{enumerate}
        \item Unique ID representing the location searched
        \item Metropolitan area of the searched location
        \item Unique id of the specific cell on the earth's surface containing the searched location's center
        \item Whether the searched location is an address, POI, street, neighborhood, or city
        \item Country of the searched location
        \item Number of guests
        \item Whether the search is from our mobile application
        \item Device type
        \item Number of lead days before trip
        \item Trip length
        \item Whether the trip is during a weekend
        \item Check in date
        \item Check out date
        \item Search date
    \end{enumerate}
    \item \textbf{Training Labels}: The latitude and longitude coordinates of the booked listing attributed to the search of the training example, defined as \textit{bookedListingLat} and \textit{bookedListingLng}.
    \item \textbf{Architecture and Training}: The architecture is a two layer neural network with hidden layers of size 256. This is shown with the red outlined components in Figure \ref{pinhead_architecture} below. The model is trained for 9 hours with a learning rate of 0.001 and batch size 800 for 30 epochs. We use Tensorflow Horovod Distributed training on 10 Nvidia Tesla A10Gs machines.
\end{itemize}
\null \quad Loss construction involves several considerations that are crucial to the product and guest experience w.r.t. relevance and recall of listings from the retrieval bound, as well as the bound's reliability. The following defines the loss for a single training example
\begin{enumerate}
    \item  \textbf{Booked Listing Location (BL) Loss}: This recall loss penalizes the model if the predicted retrieval bounds do not contain their corresponding booked listing location. This encourages the model to predict retrieval bounds that would contain most listings that anyone would book for the search
\begin{equation}
\begin{split}
\mathbf{L}(BL) = D(retrievalSwLng, bookedListingLng) \\
+ D(retrievalNeLng, bookedListingLng) \\
+ D(retrievalSwLat, bookedListingLat) \\   
+ D(retrievalNeLat, bookedListingLat),
\end{split}
\end{equation}
where D(a, b) is the geographical distance between coordinates a and b.
\item  \textbf{Retrieval Bounds Size (RBS) Loss}: This loss penalizes the model for predicting retrieval bounds that are unnecessarily large. This loss acts as a "quasi-negative" and prevents the model from predicting retrieval bounds that cover the entire world to minimize the Booked Listing Location Loss.
\begin{equation}
\begin{split}
    \mathbf{L}(RBS) = W(retrievalNeLng, retrievalSwLng) \\ 
    + H(retrievalNeLat, retrievalSwLat),
\end{split}
\end{equation}
where W(a, b) and H(a, b) are the difference in degrees between points a and b converted to kilometers
\item \textbf{Valid Bounds (VB) Loss}: This loss penalizes the model for predicting invalid bounding boxes. This occurs if the predicted southwest bound is north or east of the predicted northeast bound.
\begin{equation}
\begin{split}
        \mathbf{L}(VB) = max(retrievalSwLat - retrievalNeLat, 0) \\
        + max(retrievalSwLng - retrievalNeLng, 0),
\end{split}
\end{equation}
\end{enumerate}
\textbf{Final Model Loss}: The model is trained using a weighted combination of the losses defined above. Loss weights were tuned offline to have the same average retrieval bounds size as the previous baseline method but higher recall of booked listing locations resulting in \begin{math} \alpha = 250, \beta = 1, \gamma = 1000000 \end{math}.  This heuristic for weight tuning was used to prevent expanding the bounds too much in order to comply with product requirements. The values of each loss have very different distributions so these hyperparameters values do not directly reflect their relative importance. In practice, we only tune \begin{math}\alpha\end{math} to modify the tradeoff between booked listing location recall and the size of the bounding box while keeping \begin{math}\beta\end{math} constant. \begin{math}\gamma\end{math} is not very sensitive and we have set it arbitrarily high in order to reflect the importance of predicting bounds that can actually be used for retrieval.
\begin{equation}
\begin{split}
    \mathbf{L} = \alpha * \mathbf{L}(BL) + \beta * \mathbf{L}(RBS) + \gamma * \mathbf{L}(VB)
\end{split}
\end{equation}
\section{
\textbf{Reinforcement Learning Based Location Retrieval}}
\subsection{\textbf{The opportunity}} 
\null \quad The initial machine learning approach was cyclical in nature. We only learned booking preferences relative to search parameters within the map areas that were selected by the preexisting model. Exploring and learning for searches where the model has little confidence and updating the model with newly collected data could address this bias and improve the experience.
\subsection{\textbf{The challenge of location retrieval exploration}}
\null \quad Location retrieval can be reformulated as a reinforcement learning problem, more specifically a contextual multi-armed bandit task, in order to explore for new locations and exploit the known. The context is defined by a search with location, dates, etc. There are 4 continuous actions, defined as northern, southern, eastern and western expansion from the center of the searched location with the potential reward of a booking. \\
\null \quad Approaches for contextual multi armed bandits require the following three components
\begin{enumerate}
    \item \textbf{Active Contextual Estimation}: A method to estimate actions and rewards given a context that is regularly updated based on new exploration and feedback
    \item \textbf{Uncertainty Estimation}: A method to determine the confidence of the estimator's predictions
    \item     \textbf{Exploration Strategy}: A method to facilitate exploration and data collection for actions

\end{enumerate}
Additionally, potential solutions had to fulfill the following product constraints
\begin{enumerate}
    \item \textbf{Deterministic}: The prediction should be stable and consistent across different guests for the same query on the same day while ranking handles personalization.
    \item \textbf{Optimistic Confidence Based}: Only optimistically explore using reasonably, larger retrieval bounds for searches with low confidence predictions
    \item \textbf{Performant}: Solutions must satisfy the latency and performance requirements to serve Airbnb traffic.
\end{enumerate}
\null \quad Formulating a solution to the contextual multi-armed bandit problem for location retrieval with common methods that satisfied these constraints proved to be challenging.
\subsection{Active Contextual Estimation}
\null \quad The existing model based location retrieval solution would satisfy the requirement for a contextual estimator. Given the frequency of bookings on Airbnb, we employed daily retraining so that the estimator could be regularly updated based on new booking behavior from exploration.
\subsection{Uncertainty Estimation}
\null \quad Uncertainty estimation proved to be the most challenging to construct with the existing unusual model formulation and loss. We initially considered the following traditional methods.
\begin{enumerate}
    \item \textbf{Ridge regression} \cite{hoerl1970ridge}: regularized linear regression with least squares loss that allows for uncertainty estimation based on the standard errors of its coefficients
    \item \textbf{Bayesian deep learning} \cite{neal2012bayesian}\cite{joo2020being}: Deep neural networks where weights and biases are constructed as random variables with distributions for uncertainty estimation.
    \item \textbf{Deep ensembles} \cite{lakshminarayanan2017simple}: Ensemble of deep neural networks that are jointly use to compute the prediction mean and variance that can be used for uncertainty estimation.
    \item \textbf{Evidential Networks} \cite{sensoy2018evidential}: Utilizes learned belief functions to model the predictors uncertainty
\end{enumerate}
\null \quad These prevalent methods did not fit this use case for a variety of reasons. Ridge regression assumes a linear relationship between predictor variables and the target variable which is untrue for location retrieval. It also requires using the regularized least squares loss which is not compatible with the location retrieval objective defined above. Bayesian deep learning has slower training and inference, in addition to having twice as many parameters. Meanwhile, deep ensembles would require N times more parameters increasing code, training, and operational complexity. Finally, evidential networks would increase parameters, complexity, and latency in order to score the learned belief function for uncertainty estimation. These considerations made them unfavorable for our production use case.\\
\null \quad Given these considerations, we employ \textbf{Monte Carlo Dropout}\cite{gal2016dropout} \cite{guo2020deep} to handle uncertainty estimation. Gal et. al demonstrates that any deep neural network of arbitrary depth, non linearities, and architecture with a dropout layer is equivalent to an approximation to the probabilistic deep Gaussian process. Given an example, we can score it N unique times to generate N unique predictions which is equivalent to the score distribution for the example. This score distribution can be used for uncertainty estimation. This approach requires minimal modification to the existing model architecture and any additional overhead for training and serving can be minimized through parallelization unlike some of the other methods.

\subsection{Exploration Strategy}
\null \quad There were three exploration strategies that we considered:
\begin{enumerate}
    \item \textbf{E-greedy}\cite{sutton2018reinforcement}: Greedily recommends the actions with the highest reward with probability 1-e and randomly selects other actions uniformly with probability e. 
    \item \textbf{Thompson sampling}\cite{agrawal2013thompson}: Samples from the posterior distribution of the action reward space with current knowledge; recommends the action with the highest sampled reward estimate.
    \item \textbf{Upper Confidence Bound}\cite{auer2002using}: Computes upper confidence bounds for each action using the mean estimate for an action plus a bonus based on the uncertainty of the estimate and tuneable parameter \begin{math}\lambda\end{math}. It then chooses the action with the highest upper confidence bound. 
\end{enumerate}
\null \quad Thompson sampling's random nature results in retrieval bounds that lie anywhere on the prediction distribution. This would result in different guests having different experiences for the same search and potentially egregiously small or large bounds. This violates the product constraints outlined above and made it unsuitable for location retrieval. The e-greedy exploration strategy is context agnostic and randomly decides when and how much to explore so it would not fulfill the constraints of being optimistic confidence based or deterministic. \\
\null \quad  The Upper Confidence Bound algorithm serves an optimistic, upper percentile of the prediction distribution based on confidence in its estimate. It is also stable across guests. Given these qualities, it was the most appropriate exploration strategy among the three candidates that would be applicable to location retrieval.
\subsection{\textbf{\textbf{The Approach for Location Retrieval Exploration   }}}
\begin{figure}
  \centering
  \includegraphics[width=\linewidth]{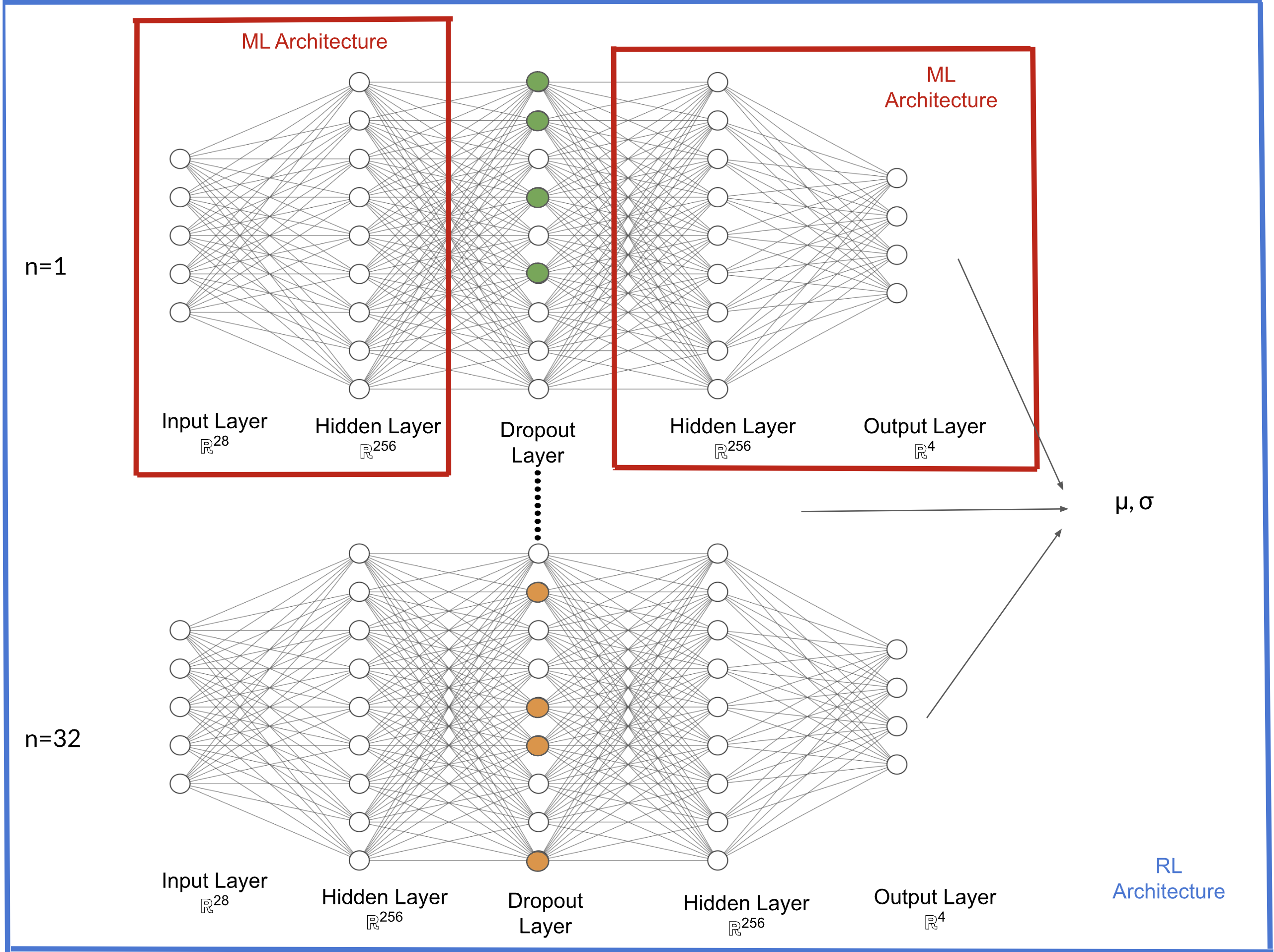}
  \caption{Location Retrieval Model Architectures}
  \footnotesize{Architecture for Section 5 in Red and Section 6 in Blue} 
  \vspace{+4mm}
  \label{pinhead_architecture}
\end{figure}

\setlength{\intextsep}{0pt}
\setlength{\textfloatsep}{0pt}
\begin{algorithm}
\caption{Exploration Enhanced Location Retrieval}
\begin{flushleft}
{\bf Input:} Search Features x \\
{\bf Output:} Location retrieval bound F = \{swLatOffsetUCB, swLngOffsetUCB, neLatOffsetUCB, neLngOffsetUCB\}
\end{flushleft}
\begin{algorithmic}
\Procedure{}{}
\State Train a DNN \begin{math} f(x) \end{math} with a random dropout layer that replaces an activation  with 0 with probability 0.95.
\State Score an example 32 times with the dropout layer enabled producing 32 unique estimates 
\begin{math} y = \{y^0, y^1, ..., y^{31}\}\end{math}, where \begin{math} y^i = f(x^i) \end{math} is the i-th estimate of the example x, represents the predicted retrieval bound offsets \{swLatOffset, swLngOffset,
         neLatOffset, neLngOffset\}
\State Compute the mean of the estimates \begin{math} \mu \end{math} for the reward estimate and the standard deviation  \begin{math} \sigma \end{math} of the estimates for uncertainty estimation.
\begin{equation} 
\mu = \dfrac{\sum_{i=0}^{31} y^i}{32},
\sigma = \dfrac{\sum_{i=0}^{31} \sqrt{(y^i - \mu)^2}}{32}
\end{equation}
\State Construct UCB retrieval bounds using \begin{math} F = \mu + 2 * \sigma \end{math} and retrieve listings in production within these bounds to facilitate exploration
\State Retrain daily based on training data derived from any new bookings to learn from exploration and increase model confidence
\EndProcedure
\end{algorithmic}
\end{algorithm}
\null \quad Algorithm 1 gives an overview of the proposed approach to solve the contextual multi armed bandit problem for location retrieval with the existing estimator, Monte Carlo Dropout, and the Upper Confidence Bound algorithm. We selected a dropout rate of 0.95 based on an offline parameter sweep to find the highest dropout rate with minimal impact to model performance. We selected N=32 for uncertainty estimation by choosing the minimum value where \begin{math} \mu \end{math} and \begin{math} \sigma \end{math} are stable across different scoring runs with the same example. We chose \begin{math} \lambda = 2 \end{math} for exploration with UCB by choosing the maximum value that still satisfied latency constraints. The model architecture consists of one hidden layer, one random dropout layer, one hidden layer and the output layer which computes the 4 coordinate offsets. This architecture is outlined in Figure \ref{pinhead_architecture}.  Model training also remains unchanged from the previous method.

\section{Experimental Results}
\subsection{\textbf{\textbf{Overview}} }
\begin{table*}
    \centering  
    \resizebox{\textwidth}{!}{%
    \begin{tabular}{l|c|c|c|c|c}
        \hline
         \textbf{Approach}&\textbf{Baseline}&\textbf{Uncancelled Bookers}&\textbf{Booked Listing Location Recall}&\textbf{Retrieval Bounds Size}&\textbf{Number of Listings Retrieved}\\
        \hline
         Heuristic 4& Heuristic 3& +0.35\%& N/A& N/A& +0.18\% \\
         Statistics& Heuristic 1, 2, 4& No Stat Sig Change&+1.05\%& N/A& -0.40\% \\
         Machine Learning& Heuristic 1, 2, 4&+1.8\%&+7.12\%&-40.83\%& +0.57\% \\
         Reinforcement Learning& Machine Learning&+0.51\%& +0.25\%&-9.3\%& +0.01\%\\
    \end{tabular}}
  \caption{Online and Offline Impact of all Approaches}
\vspace{-4mm}
  \footnotesize{Absolute metric values are not shown for company privacy reasons.}
\end{table*}
\null \quad We evaluate location retrieval models offline on held out test data. The held out data is composed of 1 weeks worth of bookings and searches from dates not included in training dataset construction equating to roughly 7 million examples. We consider the following two metrics based on the losses described above. These two metrics are typically inversely correlated, i.e., the larger, less relevant the bounds are the higher the recall of booked listing locations.
\begin{enumerate}
\vspace{-1mm}
    \item \textbf{Booked Listing Location Recall}: Percentage of booked listing locations that are contained within the predicted retrieval bounds of attributed searches
    \item \textbf{Retrieval Bounds Size}: The average size in kilometers of the predicted retrieval bounds 
    \item \textbf{Number of Listings Retrieved}: The average number of listings retrieved using the predicted bounds 
\vspace{-4mm}
\end{enumerate}
\null \quad The primary evaluation to determine whether a new approach will be launched to 100\% of production traffic is to test whether it has a statistically significant (p-val < 0.05) increase in the following business metric in a massive online A/B experiment.
\\\null \quad \textbf{Uncancelled Bookers}: The unique number of guests that complete a homes booking reservation on Airbnb that did not cancel the reservation during the experiment period. \\
\null \quad All methods were tested in 3 week 50/50 A/B experiments with \textasciitilde 54M guests worldwide. Advancements in location retrieval primarily increase uncancelled booker conversion by surfacing new listings that were not shown before that offer differing values, amenities, or style. Given Airbnb's scale and the long tail booking objective, an improvement of even just 0.2\% is very significant and difficult to achieve for optimizations within search. The impact of each method on offline and online evaluation criteria are detailed above in Table 1. Additionally, we do qualitative evaluation of specific searches, retrieval bounds, and search results to determine whether solutions test the original hypotheses. \\
\vspace{-3mm}
\subsection{\textbf{\textbf{Cold Start Heuristic based Location Retrieval}} }
\begin{figure}
  \centering
  \includegraphics[width=\linewidth]{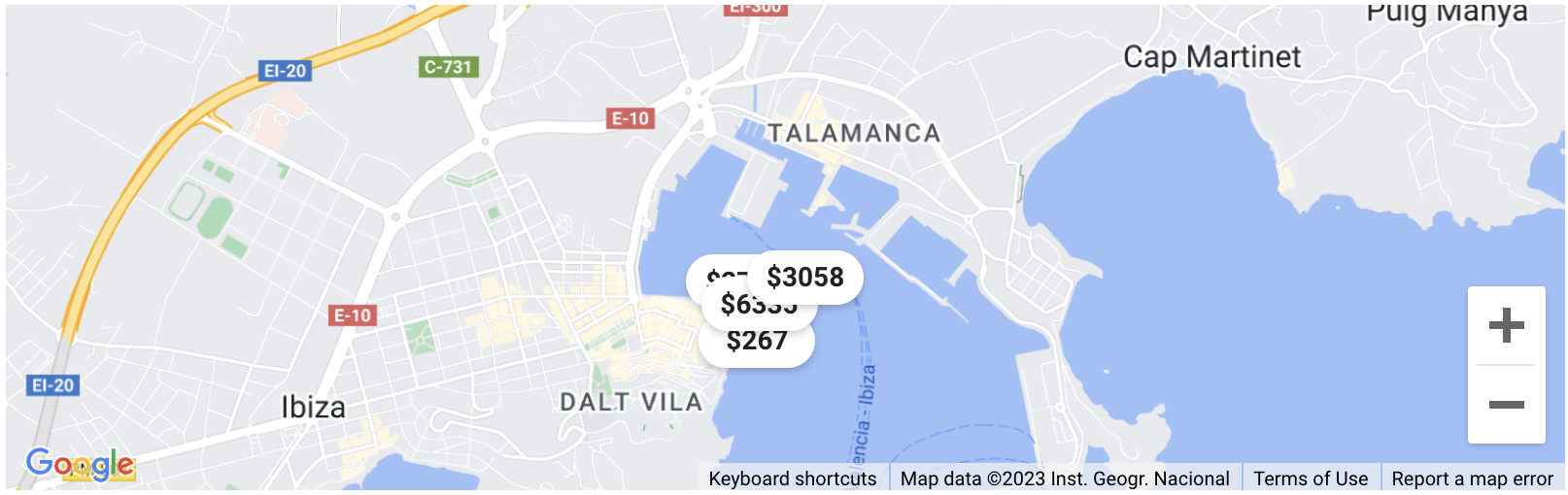}
  \caption{Search for Building in Ibiza, Spain Before Log Scale Expansion Heuristic}
  \label{ibiza_before_heuristic}
\end{figure}

\begin{figure}
  \centering
  \includegraphics[width=\linewidth]{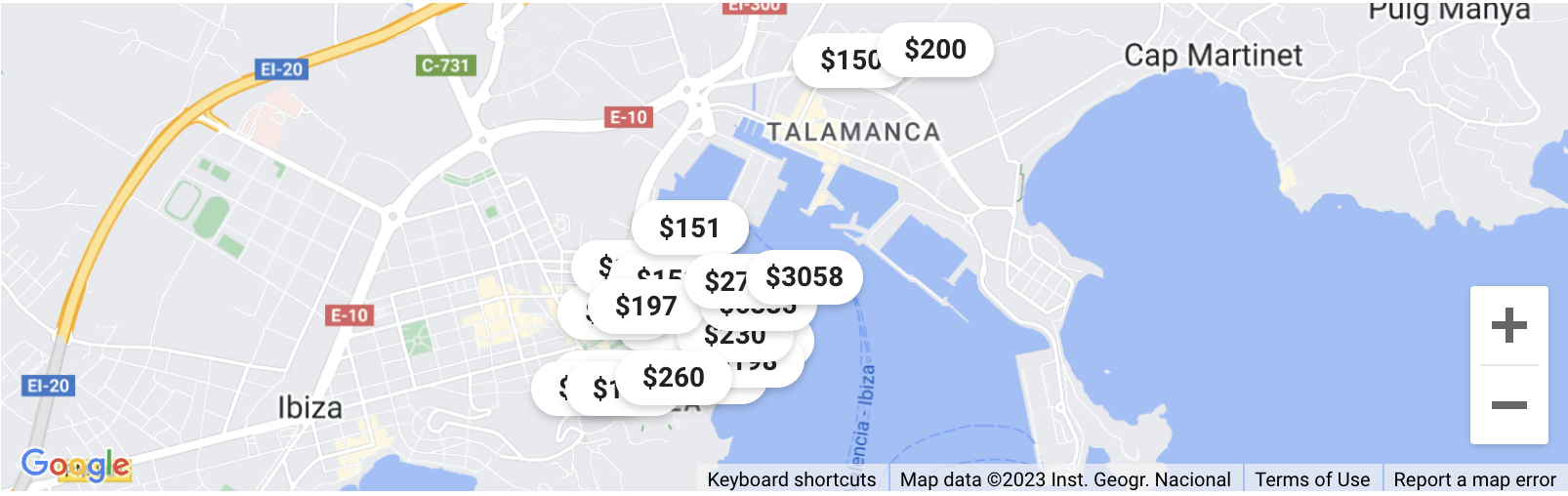}
  \caption{Search for Building in Ibiza, Spain After Log Scale Expansion Heuristic}
  \label{ibiza_after_heuristic}
\end{figure}

\null \quad The log scale expansion function (Heuristic 4) was tested in online A/B experiments against the baseline method (Heuristic 3) outlined in Section 2 for each location type. It significantly improved the launch criterion of uncancelled bookers by 0.35\% and was launched to 100\% of production traffic. Offline impact is not shown in Table 1 because it was developed and tested before our offline evaluation criteria was defined and tracked. \\
\null \quad These improvements proved that location retrieval was an impactful lever to improve experiences for guests and booking behavior. These heuristics improved searches that had disproportionately low inventory states and unbookable inventory. For example searches for a specific building in Ibiza, Spain previously resulted in 5 results with the baseline while the new heuristic surfaced several pages of bookable results by expanding the bounds less than a mile. This is demonstrated in Figures \ref{ibiza_before_heuristic} and \ref{ibiza_after_heuristic}.

\subsection{\textbf{\textbf{Statistics based Location Retrieval}} }
\null \quad The statistics based method was tested in an online A/B experiment against the three heuristics for each searched location typed defined in Section 1. However, it did not have a stat sig impact on uncancelled bookers so it was never launched to 100\% of production traffic. This experiment showed that naively taking advantage of booking data segmented only by searched location was not enough to materially improve the amount of bookers on the platform. We believed this was likely due to the methods inability to generalize across different searches and differentiate guest experiences because it was coarsely keyed by searched location. The need for this is evidenced by the difference in booking behavior between different guest counts, demonstrated in Figure \ref{sf_heatmap} and \ref{sf_heatmap_group} above.
\subsection{\textbf{\textbf{Machine Learning based Location Retrieval}} }
\begin{figure}[h]
  \centering
  \includegraphics[width=\linewidth]{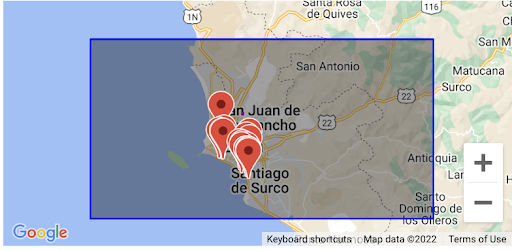}
    \caption{Retrieval Bounds and results for a search for a small street in Lima, Peru before machine learning based location retrieval}
\label{lima_before_ml}
\end{figure}

\begin{figure}[h]
  \centering
  \includegraphics[width=\linewidth]{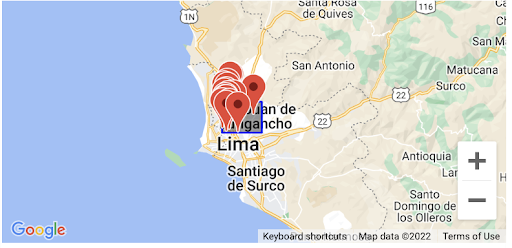}
      \caption{Retrieval Bounds and results for a search for a small street in Lima, Peru after machine learning based location retrieval}
      \label{lima_after_ml}
\end{figure}

\null \quad Applying machine learning for location retrieval was profoundly impactful. The methodology described and the following impact are the result of the initial model launch and two successive iterations on the input features and attribution. The feature engineering effort added features indicating the business market and a unique identifier for the specific cell on the Earth's surface that corresponds to the search location center in order to generalize better for rare locations. It also introduced trip dates to better handle seasonality. The attribution iteration expanded the training data to include booked listing locations that were found through map pan and zoom searches. The initial model launch was tested against the three heuristics for each searched location type defined in Section 2 and each iteration was tested successively with the previous model as a baseline for offline evaluation and online experiments. The methods cumulatively increased booked listing location recall by 7.12\% and reduced retrieval bounds sizes by 40.83\%, according to offline evaluation on held out test data compared to the baseline method.
\\ \null \quad The machine learning based location retrieval solutions were also tested successively in online A/B experiments resulting in a cumulative +1.8\% uncancelled bookers by showing guests more bookable inventory near their travel destinations. Each iteration had statistically significant positive impact to uncancelled bookers, was launched to 100\% of production traffic for a period  of time, and served as the following iteration's baseline for offline and online evaluation. 
\\ \null \quad We also qualitatively evaluated searches to verify whether the model generalizes for uncommon locations. For example, the baseline retrieval bounds are arbitrarily large with many results that were extremely far away and irrelevant for a search for a small street in Lima, Peru such that the searched street is not visible in the UI. The machine learning based retrieval bounds results in much smaller bounds and more relevant results near the street demonstrating its ability to generalize for uncommon locations. This is shown in Figures \ref{lima_before_ml} and \ref{lima_after_ml}.
\vspace{0mm}
\subsection{\textbf{\textbf{Reinforcement Learning based Location Retrieval}} }
\begin{figure}[htbp]
   \centering
   \ \caption{Mean and UCB Retrieval Bounds from Searches San Francicso, CA (left) and Smith Mountain Lake, Virginia (right) }
    \includegraphics[width=0.5\textwidth]{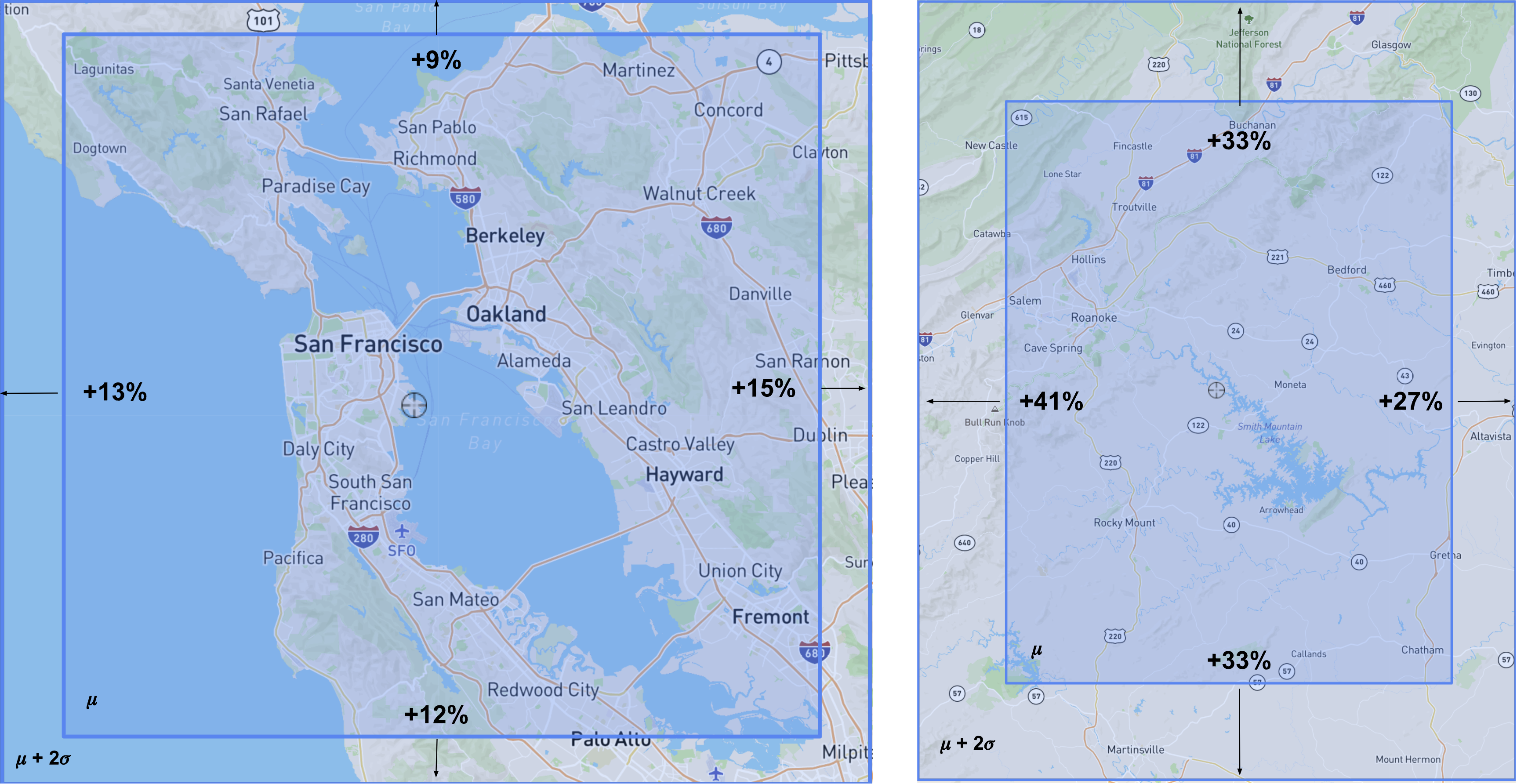}
  \label{combo_ucb_bounds}
  \end{figure}

\null \quad The MC Dropout Location Retrieval model performs better than the baseline machine learning model when evaluating offline using the mean prediction \begin{math} \mu \end{math} for each test example with +0.25\% in recall and -9.3\% in retrieval size. We also qualitatively verified whether the MC Dropout Location Retrieval model explores more for locations with little prior data versus locations that are often searched and booked. For example, the difference between the mean bounds and UCB bounds for San Francisco, California is very small. However, the difference is much larger for Smith Mountain Lake, Virginia where we have little prior data. This is demonstrated in Figure \ref{combo_ucb_bounds}.
  \\ 

\null \quad The MC Dropout Location Retrieval Model was also tested in an online A/B experiment against the baseline production machine learning model retrained on data up to the start of the experiment and the production model trained on data up to 8 months before the start of the experiment. There was no statistically significant difference in performance between the production and retrained production models. The MC Dropout model showed a statistically significant increase of +0.51\% in uncancelled bookers compared to the retrained production model, which was primarily driven by surfacing more bookable, affordable, and diverse inventory to guests. This is the result of two successive experiments. The first tested the method described against the preexisting machine learning solution. The second optimized the 32 forward passes required for uncertainty estimation by embedding it into the tensorflow graph instead of manually scoring the model 32 separate times which resulted in a significant latency improvement. Traditional multi-armed bandit regret analysis is infeasible for the nontraditional recall maximizing application but model confidence for each prediction offset has increased by 7.33\% on average since launch.

\section{Conclusion}
We presented how location retrieval was built from the ground up in Airbnb search by solving cold start, generalization, and algorithmic bias. The system evolved from heuristics all the way to the application of reinforcement learning with significant cumulative impact of +2.66\% in uncancelled bookers. This is comparable or more impactful than many previously published machine learning techniques within Airbnb search such as \cite{abdool2020managing} \cite{haldar2019applying} \cite{haldar2020improving}  \cite{tan2023optimizing} \cite{haldar2023learning}. We outlined the successes and failures along the way and the motivation for more advanced techniques to tackle the aforementioned challenges. The Monte Carlo Dropout model, the application of reinforcement learning, proved that it performed better than previous baselines offline and online for business metrics and now serves all production traffic at Airbnb. For future work, there is opportunity to advance the model’s understanding with more complex features. There is also an opportunity to reformulate the retrieval mechanism from retrieval bounds to map cells. This would allow for organically learning booking probabilities for each cell instead of training with hand tuned weights.
  
\bibliographystyle{ACM-Reference-Format}
\nocite{*}
\bibliography{base}

\end{document}